\documentclass[aps,twocolumn,superscriptaddress,showpacs]{revtex4}
\begin{document}
\def\Barcelo{Barcel\'o}
\def\Schrodinger{Schr\"odinger}
\title[Analog gravity...\dots]{Analog gravity from 
Bose--Einstein condensates}
\author{Carlos \Barcelo}
\email[]{carlos@hbar.wustl.edu}
\homepage[http://www.physics.wustl.edu/\~{}carlos]{}
\affiliation{Physics Department, Washington University, Saint Louis,
MO 63130--4899, USA}
\author{S. Liberati}
\email[]{liberati@physics.umd.edu}
\homepage[http://www2.physics.umd.edu/\~{}liberati]{}
\affiliation{International School for Advanced Studies, Via Beirut 2-4, 34014
Trieste. Italy}
\affiliation{Physics Department, University of Maryland, College Park,
MD 20742--4111, USA} 
\author{Matt Visser}
\email[]{visser@kiwi.wustl.edu}
\homepage[http://www.physics.wustl.edu/\~{}visser]{}
\affiliation{Physics Department, Washington University, Saint Louis,
MO 63130--4899, USA}
\date{7 November 2000; gr-qc/0011026; \LaTeX-ed \today}
\begin{abstract}

We analyze prospects for the use of Bose--Einstein condensates as
condensed-matter systems suitable for generating a generic ``effective
metric'', and for mimicking kinematic aspects of general relativity.
We extend the analysis due to Garay {\emph{et al}}, [gr-qc/0002015,
gr-qc/0005131].  Taking a long term view, we ask what the ultimate
limits of such a system might be. To this end, we consider a very
general version of the nonlinear {\Schrodinger} equation (with a
3-tensor position-dependent mass and arbitrary nonlinearity).  Such
equations can be used for example in discussing Bose--Einstein
condensates in heterogeneous and highly nonlinear systems. We
demonstrate that at low momenta linearized excitations of the phase of
the condensate wavefunction obey a (3+1)-dimensional d'Alembertian
equation coupling to a (3+1)-dimensional Lorentzian-signature
``effective metric'' that is generic, and depends algebraically on the
background field.  Thus at low momenta this system serves as an analog
for the curved spacetime of general relativity. In contrast at high
momenta we demonstrate how to use the eikonal approximation to extract
a well-controlled Bogoliubov-like dispersion relation, and (perhaps
unexpectedly) recover non-relativistic Newtonian physics at high
momenta. Bose--Einstein condensates appear to be an extremely
promising analog system for probing kinematic aspects of general
relativity.

\end{abstract}
\maketitle
\def\half{{1\over2}}
\def\L{{\mathcal L}}
\def\S{{\mathcal S}}
\def\d{{\mathrm{d}}}
\def\x{{\mathbf x}}
\def\v{{\mathbf v}}
\def\im{{\rm i}}
\def\etal{{\emph{et al\/}}}
\def\det{{\mathrm{det}}}
\def\tr{{\mathrm{tr}}}
\def\ie{{\emph{i.e.}}}
\def\bnabla{\mbox{\boldmath$\nabla$}}
\def\Box{\kern0.5pt{\lower0.1pt\vbox{\hrule height.5pt width 6.8pt
    \hbox{\vrule width.5pt height6pt \kern6pt \vrule width.3pt}
    \hrule height.3pt width 6.8pt} }\kern1.5pt}
\def\HRULE{{\bigskip\hrule\bigskip}}
\begin{widetext}

\section{Introduction}

Progress in understanding classical general relativity (GR) and
quantum field theory in curved spacetime (not to mention quantum
gravity itself), suffers greatly from the lack of experimental
feedback.  Moreover, direct experimental probes of many important
aspects of these theories (such as, for instance, Hawking radiation
from a black hole) appear to be very far from current (and even
foreseeable) technologies.  In this regard, the possibility of using
condensed matter systems (such as, for example, Bose--Einstein
condensates~\cite{Garay1,Garay2}) to mimic certain aspects of GR could
prove to be very important.  The basic idea of a condensed matter
analog model of GR is that the modifications to the propagation of a
field/wave due to curved spacetime can be reproduced (at least
partially) by an analog field/wave propagating in some material
background with space and time dependent properties.

The concept of a condensed matter analog model was first explored in a
little-known paper by Gordon, where he worked within the context of
the optical properties of dielectrics~\cite{Gordon}.  After lying
fallow for a considerable period, during the last decade or so this
idea has been revived and elaborated, mainly in the context of
considering the propagation of sound waves in a moving
fluid~\cite{Unruh1,Jacobson1a,Comer,Jacobson1b,Visser1,Unruh2,%
Hochberg,Visser2,Liberati1,Liberati-thesis}. Several other analog
systems have also been analyzed; see for example~\cite{Volovik,slow},
and also the mini-review by Jacobson~\cite{Jacobson:1999zk}. The
search for new analog models, and exploration of known analog models,
is ongoing~\cite{Reznik,nonlinear,oblique,CJM}.  Among these models, a
particularly promising one that will be the focus of this article is
the recent proposal of Garay, Anglin, Cirac, and Zoller based on
Bose--Einstein condensates (BEC)~\cite{Garay1,Garay2}.

Over the last few years a remarkable series of experiments on vapors of
rubidium \cite{Anderson95}, lithium \cite{Bradley95}, and sodium
\cite{Davis95} have led to a renewed interest in the phenomenon of
Bose--Einstein condensation~\cite{Bose24,Einstein24}. In these
experiments gases of alkali atoms were confined in magnetic traps and
cooled down to extremely low temperatures (of the order of fractions
of microkelvins). In order to observe the BEC the whole gas was allowed
to expand, by switching off the trapping potential, and monitored via
time-of-flight measurements made with optical methods.  The signature
of the BEC was a sharp peak in the velocity distribution for
temperatures below some critical value (see~\cite{Dalf99} for an
extensive review on the subject).

As Garay {\etal} have shown, perturbations in the phase of the
condensate wavefunction satisfy, in the low-momentum regime, an
equation equivalent to that of a massless scalar field in a curved
spacetime (the d'Alembertian equation $\Delta \phi=0$), but with the
spacetime metric replaced by an effective metric that depends on the
characteristics of the background condensate.  Present-day
experimental achievements, and the rapid development in magnetic
trapping techniques, seem to illuminate a viable path to
experimentally reproducing important general relativistic features
such as ergoregions and event horizons~\cite{Garay1, Garay2}.

In this paper we wish to explore the Bose--Einstein system in more
detail, formally extending the Garay {\etal} analysis as much as
possible. We will analyze a number of physically conceivable
extensions of the usual Gross--Pitaevskii approximation. In
particular, we will consider arbitrary non-linear interactions and
anisotropic mass-tensors, both depending explicitly on (space and
time) position.  In this scenario we will investigate how close we can
come to mimicking a ``generic'' gravitational field. (The simple
analog models based on ordinary fluid dynamics are somewhat limited in
this regard because the spatial slices are always conformally
flat~\cite{Unruh1,Visser1,Visser2,Liberati-thesis}). After including
these generalizations the only significant constraint left on the
effective metric is due to the irrotational nature of the condensate
3-velocity. Within this scenario we will then show how various quantum
corrections come into play in the whole analysis, distinguishing three
useful regimes: the quasi-classical, low-momentum, and high-momentum
regimes.

By using eikonal techniques we shall investigate the
high-momentum dispersion relation of the collective excitations of the
condensate.  In particular we shall study it with regard to the
possibility of analyzing the ``Hawking radiation'' that may be emitted
from ``horizons'' that form in this system.  We find a Bogoliubov-like
dispersion relation of the superluminal (more properly, supersonic)
type.  For situations with both an outer and an inner horizon, this
dispersion relation seems to led to exponential amplification of the
radiation flux coming from the outer horizon~\cite{Corley:1999rk}, (a
so-called black hole laser), in agreement with the unstable behaviour
of some of the solutions found by Garay {\etal}. (They also found some
metastable solutions whose origin might be related to the specific
boundary conditions of the configurations considered.)  Finally, we
discuss the way in which in the high-momentum limit one recovers the
underlying non-relativistic Newtonian physics.

\section{Bose--Einstein condensates}

In a quantum system of $N$ interacting bosons the crucial feature of a
Bose--Einstein condensate (BEC) is that it corresponds to a
configuration in which most of the bosons lie in the same
single-particle quantum state.  This system can be suitably described
in a second quantization framework by a many-body Hamiltonian of the
form~\cite{Dalf99}:
\begin{equation}
H  =  \int \! {\rm d}\x  \ {\widehat \Psi}^{\dagger} (t,\x)
\left[ - {\hbar^2 \over 2m} \nabla^2 + V_{\rm ext}(\x) \right]
{\widehat \Psi} (t,\x)
 +  {1\over 2} \int \! {\rm d}\x {\rm d}{\x}' \
{\widehat \Psi}^{\dagger} (t,\x) \; {\widehat \Psi}^{\dagger} (t,{\x}')
\; V({\x} - {\x}') \; 
{\widehat \Psi} (t,{\x}') \; {\widehat \Psi} (t,\x).
\label{eq:mbh}
\end{equation}
Here $V_{\rm ext}(\x)$ is some confining external 
potential, $V(\x-\x^{'})$ is the interatomic two-body potential 
(other possible multibody interactions are neglected at this stage),
and $m$ is the mass of the bosons undergoing condensation
(in current experiments these bosons are actually alkali atoms).
Finally, ${\widehat \Psi(t,\x)}$ is the boson field operator.

Although the quantum state for the Bose--Einstein condensate, as well
as its thermodynamic properties, can in principle be exactly computed
from (\ref{eq:mbh}), it is clear that for large ensemble of atoms this
approach can become impractical. It was Bogoliubov~\cite{Bogo47,LLP}
who first recognized that a natural ansatz for studying such a system
is the mean-field approach, which consists of separating the bosonic
field operator ${\widehat \Psi}(t,\x)$ into a classical condensate
contribution $\psi(t,\x)$ plus excitations $\widehat \varphi(t,\x)$:
\begin{equation}
  \label{eq:bogdec}
  {\widehat \Psi(t,\x)}=\psi(t,\x)+\widehat \varphi(t,\x).
\end{equation}
Here $\psi(t,x)$ is defined as the expectation value of the field
$\psi\equiv \langle \widehat{\Psi}(t,\x) \rangle$. It is sometimes
referred in the literature as the ``wave function of the
Bose--Einstein condensate''. Its modulus fixes the particle density of
the condensate, $\rho=N/V$, in such a way that
$|\psi(t,\x)|^{2}=\rho(t,\x)$.

The Bogoliubov decomposition (\ref{eq:bogdec}) is particularly useful
when the number of atoms which do not lie in the ground state of the
condensate is small.  In this case it is in fact possible to consider
the ``zeroth-order approximation'' to the Heisenberg equation given by
the many-body Hamiltonian (\ref{eq:mbh}) by just replacing ${\widehat
\Psi}(t,\x)$ with the condensate wave function
\begin{equation}
  \im\hbar \; \frac{\partial}{\partial t} \psi(t,\x)= [ \psi, H ]
  = \left[ -\frac{\hbar^2}{2m}\nabla ^2 +
  V_{\rm ext}(\x)  +\int {\rm d}{\x}' \
  \psi ^{\dagger}({\x}',t) \; V({\x}'-\x) \; 
  \psi (t,{\x}') \right] \psi (t,\x) \; .
 \label{eq:Heq}
\end{equation}
The next approximation is to assume the intermolecular interaction
term $V({\x}'-\x)$ represents a short range interactions. (In
helium-based Bose--Einstein systems this is {\emph{not}} the case and
considerably more complicated analysis is required.)  Part of the
theoretical interest in the previously cited heavy-alkali-atom
Bose--Einstein condensates is that these are extremely dilute
condensates; systems in which at low energy only binary collisions are
relevant.  This permits one to model the interaction with a
short-distance delta-like term times a unique self-coupling constant
$\lambda$ which is determined by the s-wave scattering length $a$ and
the atomic mass $m$
\begin{eqnarray}
  \label{eq:inter}
   V({\x}'-\x) &\approx& \lambda \; \delta({\x}'-\x),\\
  \label{eq:lambda}
   \lambda &=& {4\pi a \hbar^2\over m}.
\end{eqnarray}
The use of the approximate potential (\ref{eq:inter}) in equation
(\ref{eq:Heq}) leads to a standard closed-form equation for the weakly
interacting boson condensate
\begin{equation}
\label{E:tdlg}
 \im \hbar \; \frac{\partial }{\partial t} \psi(t,\x)= \left (
 - {\hbar^2\over2m}\nabla^2 
 + V_{\rm ext}(\x)
 + \lambda \; \left| \psi(t,\x) \right|^2 \right) \psi(t,\x).
\label{eq:GP}
\end{equation}
This equation, most commonly known as Gross--Pitaevskii (GP) equation,
(sometimes called the nonlinear {\Schrodinger} equation, or even the
time-dependent Landau--Ginzburg equation), can be associated to an
effective action of the form
\begin{eqnarray}
\S &=& \int \d t \; \d^3x \; \Bigg\{
\psi^* \left(  \im \hbar \; \partial_t 
+ {\hbar^2\over2m}\nabla^2 - V_{\rm ext}(\x) \right) \psi
- \half \lambda \;  \left| \psi(t,\x) \right|^4 \Bigg\},
\end{eqnarray}
which takes the name of the time-dependent Landau--Ginzburg
action. (There is some disagreement on terminology here: some authors
prefer to use the phrase ``time dependent Landau--Ginzburg equation''
for equation (\ref{E:tdlg}) only after formally discarding the
$i$. This has the net effect of replacing the {\Schrodinger} differential
operator by a diffusion operator. We always keep the $i$; all our
physics is oscillatory rather than diffusion-based.)

The Gross--Pitaevskii equation describes, in a simple and compact
form, the relevant phenomena associated with BEC. In particular, it
reproduces typical properties exhibited by superfluid systems, like
the propagation of collective excitations and the interference effects
originating from the phase of the condensate wave function.  It is
precisely from this (non-relativistic) theory that we shall now see
how an analogue of GR can be constructed, and possibly used for
experimental purposes.

\section{Generalized Gross--Pitaevskii equation}

The aim of our investigation is to explore the ultimate extent of the
ability of BEC systems to mimic general relativistic ones. In order to
do this in the most general way we shall now introduce a
generalization of the nonlinear {\Schrodinger} equation (\ref{eq:GP}).
In particular, we shall formally consider a series of generalizations
each of which is in principle allowed for systems undergoing
Bose--Einstein condensation, though they are not (yet) commonly
encountered in the experimental literature.

---(1) The first generalization we will make is to replace the quartic
$\half\lambda|\psi(t,\x)|^4$ by an arbitrary nonlinearity
$\pi(\psi^*\psi) = \pi(|\psi|^2)$. We note in particular that for
two-dimensional systems Kolomeisky {\etal} have
argued~\cite{Kolomeisky} that in many experimentally interesting cases
the nonlinearity will be cubic or even logarithmic in $|\psi|^2$.
(Although in two-dimensional systems standard Bose--Einstein
condensation does not occur it is experimentally almost certain that
``quasi-condensates'' exist, and possess collective excitations
treatable in an analogous way~\cite{Mullin}.)

---(2) The second generalization is that we will also permit the
nonlinearity function to be explicitly space and time dependent: $\pi
\to\pi(x,[\psi^*\psi])$ with $x\equiv(t,\x)$.  While intrinsic
properties of the bosons that we are trying to condense (like the
scattering length and mass) will typically not change from point to
point, we can certainly envisage a more general situation in which the
condensate is either selectively ``doped'', or perhaps (as suggested
in Garay~\etal~\cite{Garay1,Garay2}) physically constrained to move
thorough a narrow tube of varying cross-section, thereby altering its
effective properties in a position-dependent way. (We do not mean to
imply that this would be easy, or that such techniques are ``just
around the corner''; instead we are interested in seeing how far we
might ultimately be able to push this system.)

---(3) The third generalization we will make is to permit the mass to
be a 3-tensor: $m \to m_{ij}$.  Such anisotropic masses are well-known
from condensed matter physics where they are most typically
encountered in effective mass calculations for electrons immersed in a
band structure (see, for example~\cite{Ziman}). They have also been
discussed for the case of excitons (electron--hole couples held
together by Coulomb attraction) in BEC for semiconductors.  The doping
structure of the semiconductor and its anisotropies would give place
to an effective mass matrix for the paraexcitons (singlet excitons) at
least in the low momentum approximation~\cite{WLS95,MS00}.  Formally
we shall consider the possibility for anisotropic masses in the more
general context of the nonlinear {\Schrodinger} equation, regardless
of whether or not we are dealing with a band structure or even a
Bose--Einstein condensate. (If you wish, you might like to think of
this as a liquid-crystal BEC.)

---(4) The fourth generalization we will make is to also permit the
3-tensor mass to depend on position (both time and space).  Again we
might like to think of a doping gradient or similar situation.
Mathematically this has the effect that the mass matrix must be viewed
as a ``metric'' on a curved 3-dimensional space. It is convenient to
introduce an arbitrary but fixed (time and space independent) scale
$\mu$ with the dimensions of mass and then write 
\begin{equation}
m_{ij} = \mu \; {}^{(3)}h_{ij}
\end{equation}
with ${}^{(3)}h_{ij}$ being a properly dimensionless 3-metric. Note
that the introduction of $\mu$ is a {\emph{mathematical convenience}},
not a {\emph{physical necessity}}, and that all properly formulated
physical questions will be independent of $\mu$.

---(5) The fifth and last generalization will be to allow
time-dependence for the confining potential $V_{\rm ext}=V_{\rm
ext}(t,\x)$. (This is already implicit in the analysis of Garay
{\etal}, but is somewhat non-standard from the usual condensed-matter
viewpoint.)

\bigskip

We do not claim that any of these generalizations will be easy to achieve
experimentally, it is sufficient for our purposes that they are at
least physically conceivable.  After this series of generalizations we
obtain an action which now reads:
\begin{eqnarray}
\S &=& \int \d t \; \d^3x \; \sqrt{\det\left[{}^{(3)}h\right]} \; 
\Bigg\{\psi^* \bigg( \im \hbar \; \partial_t 
+ {\hbar^2\over2 \mu}\; \Delta_h 
+{\xi\hbar^2\over2\mu} \; {}^{(3)}R(h)
- V_{\rm ext}(t,\x) \bigg) \psi
-\pi(x,|\psi|^2) \Bigg\}.
\end{eqnarray}
Here $\Delta_h$ is the 3-dimensional Laplacian defined by
\begin{equation}
\Delta_h \psi = 
{1\over\sqrt{\det{\left[{}^{(3)}h\right]}}} \; 
\nabla_i \left(
\sqrt{\det{\left[{}^{(3)}h\right]}} \; 
\left[{}^{(3)}h^{-1}\right]^{ij}\;  \nabla_j \psi
\right),
\end{equation}
where $ [{}^{(3)}h^{-1}]^{ij}$ is the inverse of the 3-metric
${}^{(3)}h_{ij}$.  Additionally note the presence of the DeWitt term
\begin{equation}
{\xi\hbar^2\over2\mu}\; {}^{(3)}R(h) 
\end{equation}
involving the dimensionless parameter $\xi$ and the 3-dimensional
Ricci scalar---this term arises from operator-ordering ambiguities in
going from the ``flat space'' metric to ``curved space''(going from
position-independent $m$ to a position-varying effective
mass)~\cite{DeWitt,Schulman}.  We include the DeWitt term here for
completeness, and because it should be included as a matter of
principle, but note that it is unlikely to lead to experimentally
measurable effects.

Varying this generalized action with respect to $\psi^*$ now gives the
``generalized'' nonlinear {\Schrodinger} equation that will be of central
interest in this article:
\begin{eqnarray}
\im \hbar \; \frac{\partial}{\partial t}\psi(t,\x) &=& 
- {\hbar^2\over2\mu} \; \Delta_h \psi(t,\x) 
-{\xi\hbar^2\over2\mu} \; {}^{(3)}R(h) \; \psi(t,\x)
+ V_{\rm ext}(t,\x) \; \psi(t,\x) 
+ \pi'(\psi^* \psi) \; \psi(t,\x).
\label{E:GGPe}
\end{eqnarray}
We will now demonstrate that this non-relativistic generalized
nonlinear {\Schrodinger} equation has a (3+1)-dimensional
``effective'' Lorentzian spacetime metric hiding inside it.

\section{Madelung representation and the hydrodynamic limit}

The so-called Madelung
representation~\cite{Madelung,Takabayasi,Wong,Ghosh,Sonego} of a
{\Schrodinger} wavefunction consists of writing
\begin{equation}
\psi(t,\x) = \sqrt{\rho(t,\x)} \; \exp[-\im \theta(t,\x)/\hbar].
\end{equation}
The factor of $\hbar$ is introduced for future convenience (it
suppresses $\hbar$ as much as possible in the following equations).
This implies that $\theta$ has the dimensions of an action. The
Madelung representation is well-known in the context of the ordinary
linear {\Schrodinger} equation, and generalizes to the present
situation without difficulty.

Garay {\etal}~\cite{Garay1,Garay2} substitute the Madelung
representation into the Gross--Pitaevskii equation, and take the real
and imaginary parts. We could do the same thing here with the
generalized nonlinear {\Schrodinger} equation. Alternatively we could
insert the Madelung representation directly into the action, and vary
with respect to $\theta$ and $\rho$ to deduce Euler--Lagrange
equations. Either way, you will get two equations:

\noindent
---(1) Continuity:
\begin{equation}
\partial_t \rho + {1\over \mu} \bnabla\cdot(\rho \; \bnabla \theta) = 0.
\label{E:continuity}
\end{equation}
Here and hereafter the $\bnabla$ denotes a {\emph{covariant}}
derivative with respect to the 3-metric ${}^{(3)}h_{ij}$. If we define
a ``velocity'' (and at this stage this is a purely formal step)
\begin{equation}
(\v)^i 
\equiv {h^{ij}\; \nabla_j\theta\over\mu}  
\equiv {[m^{-1}]^{ij}\; \nabla_j\theta},
\label{E:velocity}
\end{equation}
then this ``velocity'' is actually independent of $\mu$, and equation
(\ref{E:continuity}) above is formally equivalent to the usual
equation of continuity in a curved 3-space
\begin{equation}
\frac{\partial}{\partial t} \rho + \bnabla\cdot(\rho \v) = 0.
\label{E:continuity2}
\end{equation}
---(2) Quantum analogue of the Hamilton--Jacobi equation:
\begin{equation}
\frac{\partial}{\partial t} \theta + {1\over2\mu}(\nabla \theta)^2 
+ V_{\rm ext}
+ \pi'
- {\hbar^2\over2\mu}\; 
\left( {\Delta_h\sqrt\rho\over\sqrt\rho} + \xi\; {}^{(3)}R(h) \right)= 0.
\label{E:HJ}
\end{equation}
Here and hereafter $(\nabla \theta)^2$ denotes the 3-metric scalar
inner product ${}^{(3)}[h^{-1}]^{ij} \; \partial_i \theta \;
\partial_j\theta$. 

The quantity
\begin{equation}
 V_{\rm Q}(\rho,\xi) \equiv  -  {\hbar^2\over2\mu} 
 \left( {\Delta_h\sqrt\rho\over\sqrt\rho} + \xi \; {}^{(3)}R(h) \right)
\end{equation}
is a generalization (because it now includes the 3-dimensional Ricci
scalar term) of what is often called the ``quantum
potential''~\cite{Bohm,Holland}.  (Note that the quantum potential is
actually independent of $\mu$ because rescaling $\mu$ simultaneously
rescales $h_{ij}$ in a compensating manner.) With our conventions this
is now the only place where $\hbar$ appears.

In terms of the the velocity field (\ref{E:velocity}) we
can write the Hamilton--Jacobi equation in the form
\begin{equation}
\frac{\partial}{\partial t} \left[m_{ij} (\v)^j \right]
+\bnabla_i \left( {m_{pq} \; (\v)^p \; (\v)^q \over2} 
+ V_{\rm ext}
+ \pi'
+V_{\rm Q}
\right)=0.  
\label{E:HJ2}
\end{equation}
(Note that this form makes the $\mu$-independence of the physics
manifest.)  The two real equations (\ref{E:continuity}) and
(\ref{E:HJ}) [or alternatively equations (\ref{E:continuity2}) and
(\ref{E:HJ2}), subject to the definition (\ref{E:velocity})] are
completely equivalent to the generalized nonlinear {\Schrodinger}
equation (\ref{E:GGPe}).

An interesting physical regime is that in which one can safely neglect
the quantum potential term. This approximation can be justified either
as the classical limit of the theory (it corresponds to neglecting all
terms with powers of $\hbar$) or as the regime of strong repulsive
interaction among atoms. In the latter case the density profile can be
safely considered smooth and hence it is reasonable to neglect the
kinetic-pressure term $\Delta_h\sqrt\rho/\sqrt\rho$.  This, together
with the smallness of the DeWitt term, permits one to discard $V_{\rm
Q}$.

In any case one can see that neglecting $V_{\rm Q}$, the equations
(\ref{E:continuity2}) and (\ref{E:HJ2}) have the form typical of those
for superfluids in the $T\to 0$ limit. In particular we can see, by
the absence of any term proportional to $\v\times(\bnabla\times\v)$ in
equation (\ref{E:HJ2}), that the equations we are working with are
automatically vorticity free. This is generally a necessary assumption
in order to obtain tractable equations in the analog of GR from
standard hydrodynamics~\cite{Visser1,Visser2}; here it is a free
byproduct of the GP equation (and this conclusion is does not depend
on the assumption of neglecting $V_{\rm Q}$).

The hydrodynamical form of the equations (\ref{E:continuity2}) and
(\ref{E:HJ2}) allows to describe the Bose--Einstein condensate as a
gas whose pressure and density are related by the barotropic equation
of state
\begin{equation}
P(\rho) =\pi'(\rho) \; \rho - \pi(\rho) . 
\end{equation}
Therefore it is also possible to {\emph{formally}} define a local
speed of sound by varying this pressure with respect to the mass
density of the condensate ``fluid'' $\varrho=\mu\rho$
\begin{equation}
 c^2=\frac{\partial P}{\partial \varrho}=
   \frac{\pi''\rho}{\mu}.
 \label{eq:cs}
\end{equation}
We say formally because for the general anisotropic case this
``velocity of sound'' is not physical. Physically, there will be three
principal sound velocities in three orthogonal directions, and, as we
have mentioned before and will see again, the formally convenient
parameter $\mu$ does not appear in any true physical result. This does
not diminish the convenience of introducing the parameter $\mu$ for
intermediate stages of the calculation.

It is also interesting to give a more quantitative estimate of the
magnitude of this sound velocity for the Bose--Einstein systems based
on alkali atoms.  In particular we can consider the specific case of
the rubidium gas with trivial mass tensor ($\mu=m$) and standard
interaction term $\pi(t,\x,|\psi|^2)= \pi(t,\x,\rho)= {1 \over
2}\lambda \rho^2$, so that $\pi'(t,\x,|\psi|^2)=\lambda |\psi(t,\x)|^2
= \lambda \rho$.  [Here $\lambda$ is given by equation
(\ref{eq:lambda}).] In this case the equation of state becomes $ P={1
\over 2}\rho^2 $ and the speed of sound takes the well-known form
\begin{equation}
  \label{eq:csRb}
   c=\sqrt{\frac{\lambda \rho}{m}}=\frac{2\hbar}{m}\sqrt{a \rho\pi}.
\end{equation}
For the rubidium gas one has $a(^{87}\/{\rm Rb})\approx 5.77$ nm,
$m(^{87}\/{\rm Rb})\approx 86.9$ u and in standard BEC experiments
$\rho\approx 10^{15}\; {\rm cm}^{-3}$~\cite{Dalf99}. These numbers
lead to a value of the speed of sound $c\approx 6.2 \times 10^{-3}$
m/s $\approx 6$ mm/s.  This is indeed one of the lowest speeds of
sound one can experimentally obtain.  We shall see in what follows how
this number can play an important role in the simulation of
gravitational phenomena in BEC systems.

\section{Fluctuations}

Now that we have seen how it is possible, at least in some appropriate
regime, to introduce a hydrodynamical interpretation of the condensate
equations, and how a speed of sound can be meaningfully introduced, we
are naturally lead to investigate the propagation of fluctuations in
the condensate.

In order to purse such an investigation we shall linearize the
equations of motion (\ref{E:continuity}) and (\ref{E:HJ}) around some
assumed background $(\rho_0,\theta_0)$. In particular we shall set
$\rho = \rho_0 + \epsilon \rho_1 + O(\epsilon^2)$ and $\theta =
\theta_0 + \epsilon \theta_1 + O(\epsilon^2)$. Then, we will be left
with two equations for the background configuration plus two more
(often called Bogoliubov equations) for the linearized quantities.
Linearizing the continuity equation results in the pair of equations  
\begin{eqnarray}
&&
\partial_t \rho_0 + 
{1\over \mu} \bnabla\cdot(\rho_0 \; \bnabla \theta_0) = 0,
\\
&&
\partial_t \rho_1 + 
{1\over \mu} \bnabla\cdot\left(
\rho_1 \; \bnabla \theta_0 + \rho_0 \; \bnabla \theta_1
\right) = 0.
\end{eqnarray}
Here and hereafter all inner products ($a\cdot b$) are all calculated
using the 3-metric ($h_{ij}\;a^i\;b^j$).

Linearizing the Hamilton--Jacobi equation we obtain the pair
\begin{eqnarray}
&&\partial_t \theta_0 + {1\over2\mu} (\nabla\theta_0)^2 
+ V_{\rm ext} + \pi'(\rho_0) + V_{\rm Q}(\rho_0)  = 0,
\\
&&\nonumber\\
&&\partial_t \theta_1  
+  {1\over \mu} \bnabla \theta_0 \cdot \bnabla\theta_1 
+ \pi''(\rho_0) \; \rho_1 - {\hbar^2\over2\mu}\; D_2 \rho_1 = 0.
\end{eqnarray}
Here $D_2$ represents a relatively messy second-order differential
operator obtained from linearizing the quantum potential, explicitly:
\begin{eqnarray}
D_2 \rho_1 
&\equiv&
-\half  \rho_0^{-3/2} \;[\Delta_h (\rho_0^{+1/2})]\;  \rho_1
+\half  \rho_0^{-1/2} \;\Delta_h (\rho_0^{-1/2} \rho_1).
\end{eqnarray}
The linearized Hamilton--Jacobi equation may be rearranged to yield
\begin{eqnarray}
\rho_1 &=&  -
\left[  \pi''(\rho_0) - {\hbar^2\over2\mu}\; D_2 \right]^{-1}
\left(
\partial_t \theta_1 + {1\over\mu}\bnabla\theta_0 \cdot \bnabla\theta_1
\right).
\label{E:linear-euler}
\end{eqnarray}
Now substitute this consequence of the linearized Hamilton--Jacobi
equation back into the linearized equation of continuity. We obtain,
up to an overall sign, the wave-like equation:
\begin{eqnarray}
&-& 
\partial_t  
\left\{ 
\left[  \pi''(\rho_0) - {\hbar^2\over2\mu}\; D_2 \right]^{-1}
\left(
\partial_t \theta_1 + {1\over\mu}\bnabla\theta_0 \cdot \bnabla\theta_1
\right)    
\right\}
\nonumber\\
&+&
{1\over\mu} \bnabla \cdot 
\left( 
\rho_0 \; \bnabla\theta_1 
- \bnabla \theta_0 \;
\left\{ 
\left[  \pi''(\rho_0) - {\hbar^2\over2\mu}\; D_2 \right]^{-1}
\left(
\partial_t \theta_1 + {1\over\mu}\bnabla\theta_0 \cdot \bnabla\theta_1
\right)    
\right\}
\right)
= 0.
\label{E:wave-like-physical}
\end{eqnarray}
%
This wave-like equation describes the propagation of the linearized
{\Schrodinger} phase $\theta_1$. [The coefficients of this wave-like
equation depend on the background field ($\rho_0$,$\theta_0$) that you
are linearizing around.] Once $\theta_1$ is determined, then equation
(\ref{E:linear-euler}) determines $\rho_1$.  Thus this wave equation
completely determines the propagation of linearized disturbances.  The
background fields $\rho_0$ and $\theta_0$, which appear as
time-dependent and position-dependent coefficients in this wave
equation, are constrained to solve our generalized nonlinear
{\Schrodinger} equation.  Apart from this constraint, they are
otherwise permitted to have {\em arbitrary} temporal and spatial
dependencies. To simplify things construct the symmetric $4\times4$
matrix
\begin{equation}
f^{\mu\nu}(t,\x) \equiv 
\left[
\matrix{f^{00}&\vdots&f^{0j}\cr
        \cdots\cdots&\cdot&\cdots\cdots\cdots\cdots\cr
        f^{i0}&\vdots&f^{ij}\cr } 
\right].
\label{E:explicit}             
\end{equation}
(Greek indices run from $0$--$3$, while Roman indices run from
$1$--$3$.)  Then, introducing $(3+1)$--dimensional space-time
coordinates --- $x^\mu \equiv (t; x^i)$ --- the above wave equation
(\ref{E:wave-like-physical}) is easily rewritten as
\begin{equation}
\label{E:compact}
\partial_\mu ( f^{\mu\nu} \; \partial_\nu \theta_1) = 0.
\end{equation}
Here
\begin{eqnarray}
f^{00} &=& - \left[  \pi''(\rho_0) - {\hbar^2\over2\mu}\; D_2 \right]^{-1}
\\
f^{0j} &=& -\left[  \pi''(\rho_0) - {\hbar^2\over2\mu}\; D_2 \right]^{-1}\; 
{h^{jk}\;\nabla_k \theta_0\over\mu}
\\
f^{i0} &=& - {h^{ik}\;\nabla_k \theta_0\over\mu} \; 
\left[  \pi''(\rho_0) - {\hbar^2\over2\mu}\; D_2 \right]^{-1}
\\
f^{ij} &=& {\rho_0 \; {}^{(3)}h^{ij}\over\mu} -  
{h^{ik}\; \nabla_k \theta_0\over\mu} \; 
\left[  \pi''(\rho_0) - {\hbar^2\over2\mu}\; D_2 \right]^{-1}\; 
{h^{jl}\;\nabla_l \theta_0\over\mu}.
\end{eqnarray}
Thus $f^{\mu\nu}$ is a $4\times4$ matrix of {\emph{differential
operators}} (the differential operators in question consistently
operating on {\emph{everything}} to the right).  Note that the precise
placement of the $h^{ij}$ above is immaterial since the operator $D_2$
is built using $\nabla$, the covariant derivative associated with
$h^{ij}$. This remarkably compact formulation (\ref{E:compact}) is
completely equivalent to equation (\ref{E:wave-like-physical}) and is
a much more promising stepping-stone for further manipulations.

The major obstruction to interpreting this wave equation in terms of
Lorentzian geometry is the fact that $f^{\mu\nu}$ is itself still a
matrix of differential operators, not functions. We now consider
several different approximations (valid in different regimes) which
have the effect of replacing these differential operators by
functions. After making those approximations, the remaining steps are
a straightforward application of the techniques of curved space
$(3+1)$--dimensional Lorentzian geometry. (See, for
instance~\cite{Visser1,Visser2}.)

\section{Quasi-classical approximation}

The most straightforward approximation we could make is to simply
neglect $D_2$ completely: This is actually what is done in the
analysis of Garay {\etal}~\cite{Garay1,Garay2}.  We (and they) justify
this approximation by pointing out that $D_2$ is always multiplied by
$\hbar^2$ and in this sense is suitably ``small''. We shall call this
the quasi-classical approximation. The wave equation then simplifies
to
\begin{eqnarray}
&-& 
\partial_t  
     \left( 
   {(\partial_t \theta_1 + [1/\mu]\bnabla\theta_0 \cdot \bnabla\theta_1)
   \over \pi''}
     \right)
+ 
{1\over\mu} \bnabla \cdot 
     \left( 
{\lambda \rho_0 \; \bnabla\theta_1 - \bnabla \theta_0 \;
 (\partial_t \theta_1 + [1/\mu] \bnabla\theta_0 \cdot \bnabla\theta_1)
  \over\pi''}   
     \right) = 0.
\label{E:wave-physical-2}
\end{eqnarray}
To simplify things algebraically, we can define, in analogy with
equation (\ref{eq:cs}), a speed $c$ as
\begin{equation}
 c^2 \equiv {\pi'' \; \rho_0 \over \mu}, 
 \label{eq:csbk}
\end{equation}
and the background ``velocity'' $\v_0$ by
\begin{equation}
(\v_0)^i = {h^{ij}\;\nabla_j \theta_0\over \mu} = [m^{-1}]^{ij}\;\nabla_j \theta_0.
\end{equation}
(Remember that $c$ is not necessarily the physical speed of sound; it
is simply a convenient parameterization; in contrast $\v_0$ really is
$\mu$-independent and physical.)  Now construct the symmetric
$4\times4$ matrix
\begin{equation}
f^{\mu\nu}(t,\x) \equiv 
{1\over \pi''} \left[ \matrix{-1&\vdots&-v_0^j\cr
               \cdots\cdots&\cdot&\cdots\cdots\cdots\cdots\cr
               -v_0^i&\vdots&( c^2\;h^{ij} - v_0^i \; v_0^j )\cr } 
\right].
\label{E:explicit-2}           
\end{equation}
This is now just a $4\times4$ matrix of numbers.


In any Lorentzian (that is, pseudo--Riemannian) manifold the curved
space scalar d'Alembertian is given in terms of the (3+1)--metric
$g_{\mu\nu}(t,\x)$ by
\begin{equation}
\Delta \theta \equiv 
{1\over\sqrt{-g}} 
\partial_\mu \left( \sqrt{-g} \; g^{\mu\nu} \; \partial_\nu \theta \right).
\end{equation}
The (3+1)--dimensional inverse metric, $g^{\mu\nu}(t,\x)$, is
pointwise the matrix inverse of $g_{\mu\nu}(t,\x)$, while $g
\equiv \det(g_{\mu\nu})$.  Thus one can rewrite the physically derived
wave equation (\ref{E:wave-physical-2}) in terms of the d'Alembertian
provided one identifies

\begin{equation}
\sqrt{-g} \; g^{\mu\nu} = f^{\mu\nu}.
\end{equation}
This implies, on the one hand 
\begin{equation}
\det(f^{\mu\nu}) = (\sqrt{-g})^4 \; g^{-1} = g.
\end{equation} 
On the other hand, from the explicit expression (\ref{E:explicit}),
expanding the determinant in minors
\begin{equation}
\det(f^{\mu\nu}) 
= 
(\pi'')^{-4} \left[(-1) \cdot (c^2 - v_0^2) - (-v_0)^2\right] \cdot
\left[c^2\right] \cdot
 \left[c^2\right]
=
- {c^6}/(\pi'')^4.
\end{equation} 
Thus
\begin{equation}
g = - {c^6}/(\pi'')^4; \qquad \sqrt{-g} = {c^3/(\pi'')^2}.
\end{equation}
We can therefore pick off the coefficients of the inverse ``condensate
metric''
\begin{equation}
g^{\mu\nu}(t,\x) \equiv 
{\pi''\over c^3}
\left[ \matrix{-1&\vdots&-v_0^j\cr
               \cdots\cdots&\cdot&\cdots\cdots\cdots\cdots\cr
               -v_0^i&\vdots&(c^2 \;h^{ij} - v_0^i v_0^j )\cr } 
\right]
=               
{\mu\over \rho_0 c}
\left[ \matrix{-1&\vdots&-v_0^j\cr
               \cdots\cdots&\cdot&\cdots\cdots\cdots\cdots\cr
               -v_0^i&\vdots&(c^2 \;h^{ij} - v_0^i v_0^j )\cr } 
\right].
\end{equation}
This is the effective Lorentzian metric seen by the perturbations of
the phase of the condensate wave function.  At this point let us
compare this metric with the acoustic metric
of~\cite{Unruh1,Jacobson1a,Comer,Jacobson1b,Visser1,Unruh2,Hochberg,%
Visser2,Liberati1}:
\begin{itemize}
\item
(1) The two metrics (acoustic and condensate) possess the same
conformal factor, (up to a physically irrelevant constant rescaling).
In view of this we will just call it the acoustic metric from now on,
keeping in the back of our minds that the relevant ``acoustics'' is
now the propagation of oscillations in the phase of the condensate
wavefunction.
\item
(2) Note that Garay {\etal} did not keep track of the conformal
factor, as it was not needed for the points they wanted to make.
\item
(3) There are now slightly different physical interpretations for $c$
and $v_0$.
\item
(4) There is already a non-flat 3-metric $h_{ij}$ present in the
analysis, even before the linearization procedure is carried out. It
is this feature that departs furthest from the previous
implementations of the notion of ``acoustic metric''.
\end{itemize}

We could now determine the metric itself simply by inverting this
$4\times4$ matrix. On the other hand, as is by now standard, it is
even easier to recognize that one has in front of one an example of
the Arnowitt--Deser--Misner split of a $(3+1)$--dimensional Lorentzian
spacetime metric into space + time, more commonly used in discussing
initial value data in Einstein's theory of gravity --- general
relativity. The (direct) acoustic metric is easily read off as
\begin{equation}
g_{\mu\nu}(t,\x) \equiv 
{c\over\pi''}
\left[ 
\matrix{-(c^2-v_0^2)&\vdots&-v_0^k \; h_{kj}\cr
        \cdots\cdots\cdots\cdots&\cdot&\cdots\cdots\cr
        -v_0^k \; h_{ki}&\vdots&h_{ij}\cr } 
\right]
=              
{\rho_0\over c \mu}
\left[ 
\matrix{-(c^2-v_0^2)&\vdots&-v_0^k \; h_{kj}\cr
        \cdots\cdots\cdots\cdots&\cdot&\cdots\cdots\cr
        -v_0^k \; h_{ki}&\vdots&h_{ij}\cr } 
\right].
\end{equation}
Equivalently, the ``acoustic interval'' can be expressed as
\begin{equation}
ds^2 \equiv g_{\mu\nu} \; dx^\mu \; dx^\nu =
{\rho_0\over c \mu} 
\left[
- c^2 dt^2 +  h_{ij} \; (dx^i - v_0^i \; dt) \; (dx^j - v_0^j \; dt )
\right].
\end{equation}
%
\def\bseries{} 
A few brief comments should be made before proceeding:
\begin{itemize}
\item
Observe that the signature of this metric is indeed $(-,+,+,+)$, as it
should be to be regarded as Lorentzian. There is an interesting
physical subtlety here: Some alkali atomic gases have a negative
scattering length, that is, there are attractive forces between atoms
physically leading to the collapse of the BEC. A negative scattering
length is formally equivalent to an imaginary speed of sound, and in
terms of the effective metric is equivalent to a Euclidean-signature
metric. That is: negative scattering length corresponds to an elliptic
differential operator, instead of the more usual hyperbolic
differential operator. In terms of general relativity, manipulating
the sign of the scattering length corresponds to building an analog
for a signature-changing spacetime.
\item
It should be emphasized that there are (at least) {\emph{three}}
distinct metrics relevant to the current discussion:
\begin{itemize}
\item
The {\em physical spacetime metric} is just the usual flat metric
of Minkowski space
\begin{equation}
\eta_{\mu\nu} \equiv 
({\rm diag}[-c_{\rm light}^2,1,1,1])_{\mu\nu}. 
\end{equation}
(Here $c_{\rm light} = \hbox{speed of light}$.) The quantum field couples
only to the physical metric $\eta_{\mu\nu}$. In fact the quantum field
is completely non--relativistic --- $||v_0|| \ll c_{\rm light}$.
\item
The ``mass metric'' $m_{ij} \equiv \mu \; h_{ij}$ describing the
position-dependent effective mass for the fundamental particles
described by the generalized nonlinear {\Schrodinger} equation.
\item
Fluctuations in the condensate field on the other hand, do not ``see''
the physical metric at all. Perturbations couple only to the {\em
acoustic metric} $g_{\mu\nu}$.
\item
As is common to all two-metric theories, one could also construct a
distinct ``associated metric'' by the prescription
\begin{equation}
[g_{\mathrm{associated}}]_{\mu\nu} \equiv 
\eta_{\mu\sigma} \; [g^{-1}]^{\sigma\rho} \; \eta_{\rho\nu}. 
\end{equation}
There seems to not be any clean physical interpretation for this
object.
\end{itemize}
\item
The geometry determined by the acoustic metric inherits some key
properties (such as, for example, stable causality) from the existence
of the underlying flat physical metric. (See~\cite{Visser1,Visser2}.)
\item
This acoustic metric is now sufficiently general to be able to mimic a
wide class of ``generic'' gravitational fields---the presence of the
position-dependent 3-tensor mass matrix is crucial to this
observation.  In Garay {\etal}~\cite{Garay1,Garay2} the mass was both
isotropic and position-independent, consequently the spatial slices of
their analog spacetimes were always conformally flat (the same
phenomenon occurs in the acoustic geometries
of~\cite{Unruh1,Jacobson1a,Comer,Jacobson1b,Visser1,Unruh2,%
Hochberg,Visser2,Liberati1,Liberati-thesis}).  This is the fundamental
reason we have gone to the technical trouble of adding a
position-dependent 3-tensor mass. The only significant restriction on
our version of the effective metric is that the three-velocity is
irrotational (zero vorticity; curl-free).
\item
The major differences with the Garay {\etal} analysis is that their
nonlinearity was strictly quartic, their mass both isotropic and
position-independent, and that they further approximated the current
quasi-classical approximation by going to a ``geometrical optics''
version thereof that permitted them to also neglect the conformal
factor.
\item
As we already commented, one might be a little worried that the
``speed'' $c$ depends on the arbitrary but fixed scale $\mu$. But the
3-metric $h_{ij}$ also depends on $\mu$ and
\begin{equation}
c^2\; h^{ij} = \pi'' \;
\rho_0 \; [m^{-1}]^{ij}
\end{equation}
is $\mu$-independent. Similarly, $(v_0)^i$ is independent of $\mu$. If
you ask physical questions like (for instance) ``what are the null
curves of the metric $g_{\mu\nu}$?''  they are determined by the
equation
\begin{equation}
h_{ij} 
\left( {\d x^i\over\d t} - v_0^i \right)  
\left( {\d x^j\over\d t} - v_0^j \right) = c^2.
\end{equation}
And this equation is completely independent of the arbitrary fixed
scale $\mu$. Equivalently
\begin{equation}
m_{ij} 
\left( {\d x^i\over\d t} - v_0^i \right)  
\left( {\d x^j\over\d t} - v_0^j \right) = \pi'' \rho_0.
\end{equation}
\item
We should add that in Einstein gravity the spacetime metric is related
to the distribution of matter by the non-linear Einstein--Hilbert
differential equations. In contrast, in the present context, the
acoustic metric is related to the background wavefunction in a simple
algebraic fashion. There are certainly constraints on the acoustic
metric, but they arise from the generalized nonlinear {\Schrodinger}
equation; not from the Einstein equations of general relativity.
(We belabour this trivial point because we have seen it lead to
considerable confusion.)
\item
Finally we reiterate the main reason the relativity community is
interested in these systems: regions where the speed of the condensate
flow exceeds the speed of sound very closely mimic the key kinematic
features of black hole physics. We will not repeat the relevant
details as they are more than adequately dealt with elsewhere in the
literature. (See, for example,~\cite{Unruh1,Jacobson1a,Comer,%
Jacobson1b,Visser1,Unruh2,Hochberg,Visser2,Liberati1,Liberati-thesis,%
Garay1,Garay2}.)
\end{itemize}
Many features of this acoustic metric survive beyond the current
quasi-classical approximation, and we now initiate a systematic
analysis of how much further the model can be pushed.

\section{Low-momentum approximation}

The low-momentum approximation is subtly different from the naive
quasi-classical approximation. We shall now retain the $\hbar^2 \;
D_2$ term, but take the approximation that within this quantum
potential term the gradients of the background are more important than
gradients of the fluctuation. (We justify this with the observation
that gradients of the fluctuation are doubly small, being suppressed
by both a factor of $\hbar$ and a factor of the linearization
parameter $\epsilon$). Specifically we take
\begin{eqnarray}
D_2 \rho_1 
&\equiv&
-\half  \rho_0^{-3/2} \;[\Delta_h (\rho_0^{+1/2})]\;  \rho_1
+\half  \rho_0^{-1/2} \;\Delta_h (\rho_0^{-1/2} \rho_1)
\\
&\approx& 
\left\{-\half  \rho_0^{-3/2} \;[\Delta_h (\rho_0^{+1/2})]
+\half  \rho_0^{-1/2} \;[\Delta_h (\rho_0^{-1/2})]\right\}  \rho_1
\\
&=&
-\half \left\{ 
{\Delta_h\rho_0\over \rho_0^2} 
- {(\nabla\rho_0)^2\over \rho_0^3} 
\right\} \rho_1.
\end{eqnarray}
That is, under this low-momentum approximation we can simply replace
the {\emph{operator}} $D_2$ by the {\emph{function}}
\begin{equation}
D_2 \to -\half \left\{ 
{\Delta_h \rho_0\over \rho_0^2} 
- {(\nabla\rho_0)^2\over \rho_0^3}.
\right\}
\end{equation}
The net result of this approximation is that wherever the quantity
$\pi''$ appears in the naive quasi-classical analysis it should be
replaced by
\begin{equation}
\pi'' \to \pi'' + {\hbar^2\over 4\mu} \left\{ 
{\Delta_h \rho_0\over \rho_0^2} 
- {(\nabla\rho_0)^2\over \rho_0^3} 
\right\}
\end{equation}
This does not affect the background flow velocity $v_0$, but it does
modify the propagation speed so that now
\begin{equation}
c^2 \to 
{\rho_0\over\mu} \left[
 \pi'' + {\hbar^2\over4\mu} \left\{ 
{\Delta_h \rho_0\over \rho_0^2} 
- {(\nabla\rho_0)^2\over \rho_0^3} 
\right\}
\right]
=
c_{\mathrm{quasiclassical}}^2
 \left[
 1 + {\hbar^2\over4\mu\pi''} \left\{ 
{\Delta_h \rho_0\over \rho_0^2} 
- {(\nabla\rho_0)^2\over \rho_0^3} 
\right\}
\right].
\end{equation}
This is a new higher-order correction to the quasi-classical speed of
sound we have previously introduced in equation (\ref{eq:cs}); in this
sense is a generalization of the previously known results regarding
the propagation of collective excitations in BEC.  This effect was not
contemplated in the Garay {\etal} analysis.  (Justifiably so, since
they were only interested in the geometrical ``optics'' approximation
within the quasi-classical limit.) If, using the speed of sound, we
introduce a notion of ``acoustic Compton wavelength''
\begin{equation}
\lambda_c \equiv {h\over\mu c},
\end{equation}
then
\begin{equation}
\lambda_c^2={h^2\over\mu \rho_0 \; \pi''}.
\end{equation}
So this modification to the speed of sound is seen to be the first
order term in a gradient expansion governed by the dimensionless
parameter
\begin{equation}
\lambda_c {||\nabla \rho_0 || \over 4\pi \rho_0}.
\end{equation}
%

\section{Eikonal approximation}

In contrast to the low-momentum approximation, the {\emph{eikonal}}
approximation is a high-momentum approximation where the phase
fluctuation $\theta_1$ is itself treated as a slowly-varying amplitude
times a rapidly varying phase. This phase will be taken to be the same
for both $\rho_1$ and $\theta_1$ fluctuations. In fact, if one
discards the unphysical possibility that the respective phases differ
by a time varying quantity, any time-constant difference can be safely
reabsorbed in the definition of the (complex) amplitudes.

Specifically, we shall write
\begin{eqnarray}
\theta_1(t,\x) &=& 
{\rm Re}\left\{ {\mathcal A}_\theta \; \exp(-i\phi) \right\},\\
\rho_1(t,\x)   &=& 
{\rm Re}\left\{ {\mathcal A}_\rho \; \exp(-i\phi) \right\}.
\end{eqnarray}
As a consequence of our starting assumptions, gradients of the
amplitude, and gradients of the background fields, are systematically
ignored relative to gradients of $\phi$.  [Warning: what we are doing
here is not quite a ``standard'' eikonal approximation, in the sense
that it is not applied directly on the fluctuations of the field
$\psi(t,\x)$ but separately on their amplitudes and phases $\rho_{1}$
and $\phi_{1}$.]  We adopt the notation
\begin{equation}
\omega = {\partial\phi\over\partial t}; 
\qquad 
k_i = \nabla_i \phi.
\end{equation}
Then the operator $D_2$ can be approximated as
\begin{eqnarray}
D_2 \;\rho_1 &\equiv&
-\half  \rho_0^{-3/2} \;[\Delta_h (\rho_0^{+1/2})]\;  \rho_1
+\half  \rho_0^{-1/2} \;\Delta_h (\rho_0^{-1/2} \rho_1)
\\
&\approx&
+\half  \rho_0^{-1} \;[\Delta_h \rho_1]
\\
&=&
-\half  \rho_0^{-1} \;k^2 \;\rho_1.
\end{eqnarray}
A similar result holds for $D_2$ acting on $\theta_1$.  That is, under
the eikonal approximation we effectively replace the {\emph{operator}}
$D_2$ by the {\emph{function}}
\begin{equation}
D_2 \to -\half \rho_0^{-1} k^2.
\end{equation}
For the matrix $f^{\mu\nu}$ this effectively results in the
replacement
\begin{eqnarray}
f^{00} &\to& - \left[  \pi''(\rho_0) + {\hbar^2 \; k^2\over4\mu\;\rho_0} \right]^{-1}
\\
f^{0j} &\to& -\left[  \pi''(\rho_0)  + {\hbar^2 \; k^2\over4\mu\;\rho_0}\right]^{-1}\; 
{h^{jk}\;\nabla_k \theta_0\over\mu}
\\
f^{i0} &\to& - {h^{ik}\;\nabla_k \theta_0\over\mu} \; 
\left[  \pi''(\rho_0)  + {\hbar^2 \; k^2\over4\mu\;\rho_0} \right]^{-1}
\\
f^{ij} &\to& {\rho_0 \; {}^{(3)}h^{ij}\over\mu} -  
{h^{ik}\; \nabla_k \theta_0\over\mu} \; 
\left[  \pi''(\rho_0)  + {\hbar^2 \; k^2\over4\mu\;\rho_0} \right]^{-1}\; 
{h^{jl}\;\nabla_l \theta_0\over\mu}
\end{eqnarray}
(As desired, this has the net effect of making $f^{\mu\nu}$ a matrix
of numbers, not operators.)

The physical wave equation (\ref{E:wave-like-physical}) now becomes a
nonlinear dispersion relation
\begin{equation}
f^{00} \;\omega^2 + (f^{0i} +f^{i0}) \;\omega \;k_i + f^{ij} \;k_i \;k_j 
= 0.
\end{equation}
After substituting the approximate $D_2$ into this dispersion relation
and rearranging, we see (remember: $k^2 = ||k||^2 = [h^{-1}]^{ij}
\;k_i \;k_j$)
\begin{equation}
-\omega^2 + 2 \; v_0^i \; \omega k_i +
{\rho_0 k^2\over\mu}\left[\pi''+{\hbar^2\over4\mu\rho_0} k^2\right]
-  (v_0^i  \; k_i)^2  = 0.
\end{equation}
That is:
\begin{equation}
\left(\omega - v_0^i \; k_i\right)^2 =
{\rho_0 k^2\over\mu}\left[\pi''+{\hbar^2\over4\mu\rho_0} k^2\right]
\end{equation}
Alternatively
\begin{equation}
\omega=  v_0^i \; k_i  \pm
\sqrt{
{\rho_0 k^2\over\mu}\left[\pi''+{\hbar^2\over4\mu\rho_0} k^2\right]
} =  v_0^i \; k_i  \pm
\sqrt{
\rho_0\pi'' (k_i [m^{-1}]^{ij} k_j) +
\left( {\hbar \over 2 }  (k_i [m^{-1}]^{ij} k_j)  \right)^2}.
\label{eq:disprel0}
\end{equation}

In the case of an isotropic mass matrix, $[m^{-1}]^{ij}\rightarrow
1/m$. It also makes sense to set $\mu \to m$, and $h_{ij} \to
\delta_{ij}$, in which case $c$ really does represent the physical
speed of sound. Then we can write the dispersion relation in a more
illuminating form
\begin{equation}
\omega= v_0^i \; k_i  \pm
\sqrt{c^2 k^2+\left({\hbar \over 2 m}\;k^2\right)^2}.
\label{eq:disprel}
\end{equation}
Notice how the previous anisotropic dispersion relation differs from
this isotropic one: in that case there are three different physical
sound velocities in three orthogonal principal directions.

At this stage some observations are in order:

---(1) It is interesting to recognize that the dispersion relation
(\ref{eq:disprel}) is exactly in agreement with that found in 1947 by
Bogoliubov~\cite{Bogo47,LLP} for the collective excitations of a
homogeneous Bose gas in the limit $T\to 0$ (almost complete
condensation). In his derivation Bogoliubov applied a diagonalization
procedure for the Hamiltonian describing the system of bosons.

---(2) It is easy to see that (\ref{eq:disprel0}), and its isotropic
partner (\ref{eq:disprel}) actually interpolates between two different
regimes depending on the value of the wavelength $\lambda= 2\pi /
||k||$ with respect to the ``acoustic Compton wavelength''
$\lambda_c=h/(\mu c)$. (Remember that $c$ is the speed of sound; this
is not a standard particle physics Compton wavelength. Furthermore
$||k||$ is defined using the inverse 3-metric $h^{ij}$.)  In
particular, if we assume $v_{0}=0$ (no background velocity), then for
large wavelengths $\lambda \gg \lambda_c$ one gets a standard phonon
dispersion relation $\omega \approx c ||k||$. As stressed by
Braaten~\cite{Braaten98} this can be related to the fact that the
quantum theory we are working with has a $U(1)$ symmetry which is
spontaneously broken. At low momenta we are just seeing the dispersion
relation of the corresponding Goldstone mode. For wavelengths $\lambda
\ll \lambda_c$ the quasi-particle energy tends to the kinetic energy
of an individual gas particle and in fact $\omega \approx \hbar^2
k^2/(2 m)$.

---(3) The dispersion relation (\ref{eq:disprel}) exhibits a
contribution due to the background flow $ v_0^i \; k_i$, plus a
quartic dispersion at high momenta. The group velocity is
\begin{equation}
v_g^i = {\partial\omega\over\partial k_i} =  v_0^i  
\pm
{ \left(c^2+{\hbar^2 \over 2 m^2}k^2\right) 
\over 
\sqrt{c^2 k^2+\left({\hbar \over 2 m}\;k^2\right)^2} }
\; k^i
\end{equation}
Dispersion relations of this type (but in most cases with the sign of
the quartic term reversed) have been used by Corley and Jacobson in
analyzing the issue of trans--Planckian modes in the Hawking radiation
from general relativistic black
holes~\cite{Jacobson1a,Jacobson1b,CJM}.  In their analysis the group
velocity reverses its sign for large momenta. (Unruh's analysis of
this problem used a slightly different toy model in which the
dispersion relation saturated at high momentum~\cite{Unruh2}.)  In our
case, however, the group velocity grows without bound allowing
high-momentum modes to escape from behind the ``horizon''. (Thus the
acoustic horizon is not ``absolute'' in these models, but is instead
frequency dependent, a phenomenon that is common once non-trivial
dispersion is included.)

This type of ``superluminal'' dispersion relation has also been
analyzed by Corley and Jacobson~\cite{Corley:1999rk}.  They found that
this escape of modes from behind the horizon often leads to
self-amplified inabilities in systems possessing both an inner horizon
as well as an outer horizon, possibly causing them to disappear in an
explosion of phonons.  This is also in partial agreement with the
stability analysis performed by Garay {\etal} using the whole
Bogoliubov equations.  They found unstable solutions with the kind of
behaviour just mentioned, but they also find stability regions
(depending on the value of certain configuration parameters).  The
existence of this stable configurations might be related with the
specific boundary conditions imposed in their configurations.

Indeed, with hindsight the fact that the group velocity goes to
infinity for large $k$ was pre-ordained: After all, we started from
the generalized nonlinear {\Schrodinger} equation, and we know what
its characteristic curves are.  Like the diffusion equation the
characteristic curves of the {\Schrodinger} equation (linear or
nonlinear) move at infinite speed. If we then approximate this
generalized nonlinear {\Schrodinger} equation in any manner, for
instance by linearization, we cannot change the characteristic curves:
for any well behaved approximation technique, at high frequency and
momentum we should recover the characteristic curves of the system we
started with.  However, what we certainly do see in this analysis is a
suitably large region of momentum space for which the concept of the
effective metric both makes sense, and leads to finite propagation
speed for medium-frequency oscillations.

---(4) There is an amusing feature to the (generalized) Bogoliubov
dispersion relation which it may be worth making explicit: Consider
the dispersion relation
\begin{equation}
\omega(k) = \sqrt{ m_0^2 + k^2 + \left({k^2\over2m_\infty}\right)^2 }.
\end{equation}
(BEC condensates correspond to $m_0 = 0$, we retain this term here for
generality. We have made $c=\hbar=1$.) 
At low momenta ($k \ll m_0$) this dispersion relation
has the usual non-relativistic limit
\begin{equation}
\omega(k) = m_0 + {k^2\over2m_0} + O(k^4).
\end{equation}
At intermediate momenta ($m_0 \ll k \ll m_\infty$) this dispersion
relation has an approximately relativistic form. Finally at large momenta
($k \gg m_\infty$) the dispersion relation again (perhaps
surprisingly) recovers a non-relativistic form
\begin{equation}
\omega(k) = {k^2\over2m_\infty} + m_\infty + O(k^{-2}).
\end{equation}
This serves to drive home in a particularly simple way the point that
observing a Lorentz invariant spectrum does not guarantee that the
underlying physics is Lorentz invariant. Indeed the entire programme
of searching for analog models of general relativity generically seeks
to take some simple (physically accessible and typically
non-relativistic) model and determine if it nevertheless hides within
it some useful approximation to Lorentzian geometry.

\section{Summary and Discussion}
 
In seeking to see how far we might be able to push the BEC system as
an analog model for general relativity we have encountered a number of
intriguing features: First, we have shown that the existence of a
regime in which phase perturbations of the wave function of the BEC
(or quasi-BEC) behave as though coupled to an ``effective Lorentzian
metric'' is a generic feature, independent of the explicit form of the
non-linear terms in the {\Schrodinger} equation. The only exception
comes about when the forces exerted between atoms are attractive. In
this case the equation of motion for the phase perturbations are no
longer hyperbolic, so the whole notion of wave disappears for these
systems. (In GR language, this corresponds to a Euclidean-signature
metric.)

Second, we have seen that in contrast to the isotropic acoustic
systems considered to date, mimicking a generic gravitational field is
not a priori implausible though it would technically be very
challenging, relying as it does on the direct introduction of
anisotropies into the generalized nonlinear {\Schrodinger} equation
via a position-dependent 3-tensor effective mass.

We have also explicitly seen how the whole notion of ``effective
metric'' in these BEC systems is intrinsically an approximation valid
for certain ranges of frequency and wavenumber --- this should not
really be a surprise since even for normal acoustic systems eventually
the atomic nature of matter provides a natural cutoff. In the BEC
system we have seen that the acoustic Compton wavelength plays a
similar role: wavelengths long compared to the acoustic Compton
wavelength see a Lorentzian ``effective metric'' while wavelengths
short compared to the acoustic Compton wavelength probe the ``high
energy'' physics (which in this situation is the non-relativistic
{\Schrodinger} equation).

Finally, we mention that from the sound velocities typically
encountered in BEC systems we can make a crude estimate of the Hawking
temperature to be expected in these systems. Using the standard
estimate based on dimensional analysis~\cite{Unruh1,Visser2,Liberati1}
\begin{equation}
T \approx {\hbar \over 2\pi k_B} \; {c\over R},
\end{equation}
and choosing a value of $R\approx 10 \; \mu$ for the size of the
acoustic black hole, we would have $T\approx 10^{-9}$ K. While
extremely small this temperature is only three orders of magnitude
less than that of the BEC itself. Furthermore, as argued
in~\cite{Liberati1} this order of magnitude estimate for the Hawking
temperature is often misleadingly low --- and this fact, in addition
to the intrinsically interesting features of the effective geometry
approach, makes further investigation of these BEC systems worthwhile
--- both for the condensed matter and relativity communities.

\section*{Acknowledgements}

The research of CB was supported by the Spanish Ministry of Education
and Culture (MEC). MV is supported by the US Department of Energy.  SL
is currently supported by the US NSF; while much of this work was
carried out while SL was supported by SISSA, Trieste, Italy.  The
authors wish to thank Ted Jacobson for useful comments. CB and MV
acknowledge the feedback and discussion provided by the participants
in the Workshop on Analog Models of General Relativity, held in Rio de
Janeiro, October 2000.


\end{widetext}
\end{document}